\documentstyle[12pt]{article}
\begin{document}
\title{Complexified quantum and classical mechanics}
\author{Slobodan Prvanovi\'c \\
{\it Institute of Physics Belgrade, } \\
{\it University of Belgrade, Pregrevica 118, 11080 Belgrade,}\\
{\it Serbia}}
\date{}
\maketitle

\begin{abstract}
 The anti self-adjoint operators of coordinate and momentum are introduced and applied in discussion of tunneling through the potential barrier where imaginary
 value of momentum unavoidably appears. Tunneling through temporal barrier is proceeded in a similar way and it is shown that the quantum system can tunnel, while
 the classical systems are destroyed by such barrier. The imaginary observables of coordinate and momentum are used in a novel treatment of passage through the
 event horizon. By considering the event horizon as the border of the finite potential barrier, it is found that there is no singularity at the center of a black 
 hole. The imaginary variables of coordinate and momentum are used for the construction of a toy model of the universe. It is shown that the acceleration of the 
 expansion of the universe does not need dark energies to be explained since negative energies are present and characteristic for the imaginary world.


\end{abstract}

\section{Introduction}

 Tunneling through the potential barrier is a typical quantum phenomena and there are many articles discussing it from different angles and applying it
 in a huge variety of diverse situations [1-15]. Recently, attention are attracting the time dependent potentials, see for instance [16-19]. Usually, the
 potential depending on coordinate is multiplied by the time dependent function and this modulation of a barrier raises interesting effects in different
 areas of research. Here, instead of proceeding in this direction, we shall analyze behavior of the quantum system when it is confronted to the barrier that
 depends solely on time. Since we are interested in conceptual aspects of the tunneling through the temporal barrier, we shall discuss it in general, being explicit
 only in the most simple case of a square barrier. However, we shall compare behavior of the quantum system with the behaviour of classical system in the same situation
 and this will be done in order to underline strong difference between these two. So to say, the quantum system, by tunneling through the barrier, can survive, while
the classical one cannot survive the action of the temporal barrier.

 There are solutions of the Schr\"odinger equation for which the imaginary values of momentum appear. That is, by applying the self-adjoint operator of momentum
 on such solutions, one ends with an imaginary value. This surprising feature was discussed in [20,21]. Typical example of this are the solutions of the Schr\"odinger
 equation in the case of the square potential, when one considers the solutions in region where the quantum system is within the barrier, {\it i. e.}, during its
 supposed tunneling through the barrier. To end with an imaginary value after acting with the operator of momentum on the solution of the Schr\"odinger equation is
 strange since this operator is self-adjoint, so only real values are expected. Directly attached to these imaginary values of the momentum are negative values
 of the kinetic energy, which are equally well unacceptable from the point of view of our everyday experience. The same situation will appear for the time dependent
 barrier that we are going to consider here. But, we shall approach to this imaginary values in completely nonstandard way. Namely, they will lead us to, so to say,
 complexification of the coordinate and momentum. Just like beside the real axis one introduces the imaginary one, here we shall introduce anti self-adoint
 (skew symmetric) operators of coordinate and momentum. These new operators will have imaginary eigenvalues and they shall offer adequate and consistent description
of the tunneling through barriers.

 The formalism of the operator of time, that we have proposed in [22-26], will be used here. Let us just mention that there is a whole variety of topics and
 approaches related to the operator of time, {\it e.g.}, [27-29] and references therein. However, our approach is similar to [30], and references therein,
 and [31], and its crucial point is to treat time and energy on equal footing with coordinate and momentum. This means that the separate Hilbert space, where
 operators of time and energy act, is introduced, just as it is done for each degree of freedom in the standard formulation of quantum mechanics. By doing this,
 the Pauli's objection is avoided. Proceeding in this way, the same commutation relation that holds for the coordinate and momentum is imposed for the energy and
 time, which leads to unbounded spectrum of these operators. Finally, the Schr\"odinger equation appears as a constraint in the overall Hilbert space,
 selecting the states with non-negative energy for the standard Hamiltonians. But, after the algebra of anti self-adjoint operators of coordinate and momentum is
 introduced, the negative eigenvalues of energy will naturally appear in the formalism.

 The standard quantum mechanics, with its self-adjoint operators of coordinate and momentum and appropriate real eigenvalues, is a part of the complete theory that
 involves anti self-adjoint operators of coordinate and momentum and their imaginary eigenvalues. The last ones, together with the negative energies, find
 their natural explanation within the complexified quantum mechanics. So, not just that there is nothing wrong with the imaginary values of momentum in the case of
 tunneling, but these imaginary values are meaningful since, beside our part of the universe characterized by the real numbers, there is the other one. This part is
 characterized by the imaginary numbers. These two parts are on equal footing and they together form the complete universe.

 As the examples of the possible application of complexified mechanics, we shall propose new way of looking on passages of the quantum and classical systems through 
the event horizon during the fall in and possible escape from a black hole. We will discuss evaporation of a black hole and virtual particles that escaped from a 
black hole. On the other side, we shall apply complexified classical mechanics in designing a toy model of the universe with the accelerated expansion. In this 
context, we will analyze the concept of dark energy. 

\section{Operators of time and energy}

 As it is done for every spatial degree of freedom, a separate Hilbert space ${\cal H}_t$, where operators of time $\hat t$ and energy $\hat s$ act
 non-trivially, can be introduced. So, for the case of one degree of freedom, there are $\hat q \otimes \hat I$, $\hat p \otimes \hat I$, $\hat I
 \otimes \hat t$ and $\hat I \otimes \hat s$, acting in ${\cal H}_q \otimes {\cal H}_t$, and for these self-adjoint operators the following
commutation relations hold:
\begin{equation}
  {1\over i\hbar} [\hat q \otimes \hat I, \hat p \otimes \hat I ] = \hat I \otimes \hat I ,
\end{equation}
\begin{equation}
  {1\over i\hbar} [\hat I \otimes \hat t , \hat I \otimes \hat s ] = - \hat I \otimes \hat I .
\end{equation}
 The other commutators vanish. The operators of time $\hat t$ and energy $\hat s$ have continuous spectrum $\{ -\infty , +\infty \}$, just like the
 operators of coordinate and momentum $\hat q$ and $\hat p$. The eigenvectors of $\hat t$ are $\vert t \rangle $ for every $t\in {\bf R}$.
 In $\vert t \rangle$ representation, operator of energy is given by $i \hbar {\partial \over \partial t}$ and its eigenvectors $\vert E \rangle$,
 in the same representation, are $e^{{1\over i\hbar} E\cdot t}$ for every $E\in \bf R$. In [22] we have shown how the unbounded spectrum of the operator
 of energy is regulated by the Schr\"odinger equation. Namely, the Schr\"odinger equation, that appears as constraint in ${\cal H}_q \otimes {\cal H}_t$,
 selects the states with non-negative energy, due to the bounded from below spectrum of the standard Hamiltonians. That is, the Hamiltonian and operator of
 energy are acting in different Hilbert spaces, but there is subspace of the total Hilbert space where:
\begin{equation}
 \hat s \vert \psi \rangle  = H (\hat q, \hat p ) \vert \psi \rangle  .
\end{equation}
 The states that satisfy this equation have non-negative energy for the usually used Hamiltonians. The last equation is nothing else but the Schr\"odinger equation.
 By taking $\vert q \rangle \otimes \vert t \rangle$ representation of the previous equation, one gets the familiar form of the Schr\"odinger equation:
\begin{equation}
 i \hbar {\partial \over \partial t } \psi (q,t)  = \hat H \psi (q,t)  .
\end{equation}
 With the shorthand notation $\hat H  = H (q , -i\hbar {\partial \over \partial q } )$. In other words, the operator of energy has negative
 eigenvalues as well as non-negative, but the states with non-negative energies, that are used in the standard quantum mechanics, are selected by the
 Schr\"odinger equation due to the non-negative spectra of $H (\hat q , \hat p )$. For the time independent Hamiltonian, the typical solution of the
 Schr\"odinger equation $\psi _E (q) e^{{1\over i \hbar} E \cdot t}$ is $\vert q \rangle \otimes \vert t \rangle$ representation of
 $\vert \psi _E \rangle \otimes \vert E \rangle $, where $H (\hat q , \hat p ) \vert \psi _E \rangle = E \vert \psi _E \rangle$
 and $\hat s \vert E \rangle = E \vert E \rangle$. The energy eigenvectors $\vert E \rangle$ have the same formal characteristics as, say, the
 momentum eigenvectors (they are "normalized" to $\delta (0)$ and, for different values of energy, they are mutually orthogonal).

The negative eigenvalues of $\hat s$ will appear in discussion below.

\section{Spatial barrier}

 Let us briefly review the standard treatment of the tunneling through the square potential. If the potential vanishes everywhere except between $q_a$ and
 $q_b$, where it is constant having value $V=V_0$, one firstly solves the Schr\"odinger equation for the regions where $V=0$ and then for the region where
 $V=V_0$. If the energy of the quantum system is taken to be $E_0$, then the solutions outside the barrier, in the coordinate representation, are plain vawes
 $e^{\pm {1\over i \hbar} p \cdot q}e^{{1\over i \hbar} E_0 \cdot t}$, where $E_0 ={p^2 \over 2m}$. For the region whith $V=V_0$ one finds $e^{\pm {1\over i \hbar}
 p'\cdot q} e^{{1\over i \hbar} E\cdot t}$, where $E_0 ={p'^2 \over 2m} + V_0$. We are talking about tunneling when $E_0 < V_0$. Then, the quantum system can be
 in the region that is forbidden from the point of view of classical mechanics. Namely, the classical system, having momentum $p$, will approach the potential barrier,
 say from the left, and at the point $q=q_a$ it will be pushed back by the force caused by the barrier. It will never enter the region between $q_a$ and $q_b$
 since it does not have enough energy to overcome repulsive force that acts at $q_a$.

 Discussion regarding the tunneling through the potential barrier usually proceeds by the calculation of reflection and transition rates. Instead of repeating this, 
let us focus on the difference between classical and quantum system that is essential  for tunneling. Namely, regarding the position, the classical system is always 
located at one point. Its state, {\it i .e.}, the coordinate part of  the state, is given by $\delta (q - q(t))$. So, at some moment, the classical system starts to 
feel the presence of the barrier. More precisely, only at the moment when it comes to $q_a$ it feels the repulsive force which pushes it back to the region from which 
it came. In difference to the classical  system, the quantum system, having energy $E_0$, is not located in one point of space. Its state spreads all over the space, 
and it is given by $e^{\pm {1\over i \hbar} p\cdot q}$, for $q$ between $-\infty $ and $q_a$, $e^{\pm {1\over i \hbar} p'\cdot q}$, for $q$ between $q_a$ and $q_b$, 
and $e^{\pm {1\over i \hbar}  p\cdot q}$, for $q$ between $q_b$ and $+\infty$. In the present context, being in the state with sharp value of energy actually means 
that quantum system has sharp value of momentum, so it is not in a state that is spatially located. Since we are not measuring coordinate of the quantum system, we 
are not reducing its state to the one with sharp value of position, in particular not to the state $\vert q_a \rangle$. In this way, quantum system is never actually 
infront of the barrier, so it does not feel its repulsive force. So to say, being in a state with sharp value of energy, quantum system is potentially in-front, 
inside and beside the barrier simultaneously, and its existence is not concentrated to some particular point of space.

\section{Imaginary momentum}

 If one applies momentum operator $-i \hbar {\partial \over \partial q}$ to the states of quantum system $e^{\pm {1\over i \hbar} p'\cdot q} e^{{1\over
 i \hbar} E\cdot t}$, where $E_0 ={p'^2 \over 2m} + V_0$, then for $E_0 < V_0$ one ends with the imaginary value of momentum. This is, of course, strange
 since we do not expect to find imaginary values when the self-adjoint operators act on perfectly correct solutions of the Schr\"odinger equation. There are
 other situations for which we find imaginary momentum, as well, see [32]. (There is essential difference between the present considerations and there
 given one. Namely, here the Hamiltonian and momentum commute, while in [32] this is not the case.)

 Since the imaginary values are unavoidable in the considered cases, then their meaning should be addressed in the first place. As is well known, we use in
 quantum mechanics hermitian and self-adjoint operators because they have real eigenvalues. The world we live in is characterized by the real numbers as
 the values of considered quantities. So, if we say, in a case when some system is in a state that has real eigenvalue for some observable, that this quantity
 belongs to our real world, then in the case when the system is in a state that gives imaginary value for some observable we cannot attribute reality to this
 quantity. This quantity with imaginary value does not belong to the real world. It belongs to the world of imaginary quantities.

 The unavoidable imaginary values of the momentum for $e^{\pm {1\over i \hbar} p'\cdot q} e^{{1\over i \hbar} E\cdot t}$ 
(where $E_0 ={p'^2 \over 2m} + V_0$ and $E_0 < V_0$) are demonstration of the existence of the world other than ours. So, there are two worlds. One is the world 
characterised with properties  that are real. It is our world where everything is expressed by real numbers, but there is another world where everything is 
expressed by imaginary numbers. This world  of unreality is on equal footing with ours. It is as natural as is our world. For the sake of simplicity, let us 
call that other world the imaginary world. The universe consists of both the real and the imaginary world.

\section{Anti self-adjoint operators}

 As we have underlined, it is unavoidable to end with the imaginary values of the momentum for the above given states. Since this is not consistent with the fact that
 momentum is self-adjoint operator, the question is whether there exists operator of momentum with which we can reproduce all important features of the
 considered tunneling, being such that the imaginary values are in accordance with its nature. The answer is affirmative. Namely, there are anti self-adjoint
 operators which have imaginary eigenvalues. For such an operator it is completely consistent to end with the imaginary value after applying this operator to some
 state, just as it is consistent to get real value when a self-adjoint operator is applied to some state.

 Let us introduce anti self-adjoint operators of imaginary coordinate and imaginary momentum in the following way. For each real
 $q_{re}$ there is $\vert q_{re} \rangle$ which is the eigenvector of the self-adjoint operator of real coordinate $\hat q _{re} =
\int q_{re} \vert q_{re} \rangle \langle q_{re} \vert dq_{re}$:
\begin{equation}
 \hat q _{re} \vert q_{re} \rangle =  q_{re} \vert q_{re} \rangle .
\end{equation}
 The self-adjoint operator of the real momentum $\hat p _{re}$, in the basis $\vert q_{re} \rangle$, is represented by $-i \hbar {\partial \over \partial
 q_{re}}$. Its eigenvectors, in the same representation, are (real) plain waves $e^{{1\over i \hbar} p_{re} \cdot q_{re}}$ with the real eigenvalues. These two
 opearators, acting in the rigged Hilbert space ${\cal H}_{re}$ are the standard operators of coordinate and momentum that are used in quantum mechanics. They form
 the Heisenberg algebra and they do not commute ${1\over i \hbar}[ \hat q _{re} , \hat p _{re} ] = \hat I _{re}$, where $ \hat I _{re} = \int \vert q_{re} \rangle
\langle q_{re} \vert dq_{re} $.

 Beside this, there is the algebra of non-commuting anti self-adjoint operators of coordinate $\hat q _{im}$ and momentum $\hat p _{im}$ acting in the
 rigged Hilbert space ${\cal H}_{im}$. The spectral form of the imaginary coordinate $\hat q _{im}$, in the basis $\vert q_{im} \rangle$, where $q_{im}$
 ranges over entire imaginary axis, is $\hat q _{im} = \int q_{im} \vert q_{im} \rangle \langle q_{im} \vert dq_{im}$ (with the real measure $dq_{im}$), and its
action on the eigenvectors is given by:
\begin{equation}
 \hat q _{im} \vert q_{im} \rangle = q_{im} \vert q_{im} \rangle .
\end{equation}
 In the basis $\vert q_{im} \rangle$, the anti self-adjoint operator of imaginary momentum is represented by $-i \hbar {\partial \over \partial q_{im}}$.
 Its eigenvectors, in the same representation, are the imaginary plain vawes $e^{{1\over i \hbar} p_{im} \cdot q_{im}}$ with the imaginary eigenvalues. In similarity
 to the case of the real coordinate and momentum, the commutator of the imaginary ones is proportional to the $\hat I _{im}$ which is $\hat I _{im} = \int
\vert q_{im} \rangle \langle q_{im} \vert dq_{im} $.

 In the case of one degree of freedom, the complete description demands direct product of the rigged Hilbert spaces ${\cal H}_{re}$ and ${\cal H}_{im}$. In
 ${\cal H}_{re} \otimes {\cal H}_{im}$ the operator of real coordinate is $\hat q _{re} \otimes \hat I _{im}$, while the operator of imaginary coordinate is
 $\hat I _{re} \otimes \hat q _{im}$ (and similarly for the operators of momentum). These operators, together with the identity, form the basis of complexified
quantum mechanics.

 The quantum and classical mechanics have to be identical regarding the formal structures. Therefore, there are the imaginary variables of coordinate and momentum
 $q_{im}$ and $p_{im}$ beside the standard real $q_{re}$ and $p_{re}$. These variables will be used below.

\section{Temporal barrier}

 The opportunity to be in a superposition of states is the essential for tunneling through the spatial barrier. When the quantum system is in the state with sharp value
 of momentum it is potentially everywhere without actually being there. This is related to the noncommutativity of the operators of coordinate and momenta. On the other
 side, the classical system simultaneously poseses sharp values of coordinate and momentum since they commute. The similar situation is with the energy and time. The
 oprators $\hat s$ and $\hat t$ do not commute, so they do not have common eigenvectors. When the quantum system is in the state with the sharp value of energy, than
 it is not localized in time - it is not actually present in any moment of time. The classical system, of course, is always present and has definite energy since the
 variables $s$ and $t$ commute. As in the case of the spatial barrier, this difference between classical and quantum systems is what makes possible tunneling through 
the temporal barrier.

 How the temporal barrier would affect the classical system can be viewed in the following situation. Let the initially free classical system interacts with some other
 one in such a way that it loses its mechanical energy during time, transferring it to that other system. Within the formalism of complexified classical mechanics,
let the total energy of these systems be:
\begin{equation}
{p^2 _{re} \over 2m}+ {p^2 _{im} \over 2m} + W(t),
\end{equation}
 where $W(t)$ is the energy of the other system for which we assume that, at $t_1$, it starts to rise smoothly from zero, overcomes the initial energy of the considered
 system, and then, after some time, smoothly drops down becoming zero at $t_2$. As it rises, the energy of the classical system lowers in order to keep the total energy
 of these systems constant. At some moment $t_a$, $t_1 < t_a$, $W(t)$ becomes equal to the energy that the system had before $t_a$, so the energy of the system
 becomes equal to zero. At some $t_b$, $t_a < t_b$, $W(t)$ drops to the value of the initial energy of the system, and then at $t_2$, $t_b < t_2$, vanishes. If the
 classical system initially was in the state with $p _{re}= p ^0 _{re}$ and $p _{im}= 0$, so its all energy is in the positive kinetic energy, it is crucial to find out
what happens at $t_a$.

 In between $t_1$ and $t_a$ system loses positive kinetic energy of the motion in the real world, at $t_a$ it becomes zero and then negative since $W(t)$ rises. What
 this means can be better understood in the full relativistic aproach. So, the energy of the system is given by:
\begin{equation}
 {mc_{re} ^2 \over (1- {v_{re} ^2 \over c_{re} ^2})^{1\over 2}} + {mc_{im} ^2 \over (1- {v_{im} ^2 \over c_{im} ^2})^{1\over 2}},
\end{equation}
 where $c_{re}=c$ is the speed of light in the real world and $c_{im}$ is the speed of light in the imaginary world. (The speed of light in the imaginary world is
 imaginary, as are all velocities, and $c_{im}=i \cdot c$.) In between $t_1$ and $t_a$, due to the fourth component of the 4-vector of force in the real world, which
 is equal to $-{1\over c_{re}}{dW(t)\over dt} $, the fourth component of the 4-vector of momentum in the real world decreases until it becomes $mc_{re}$ at $t_a$. Since
 $W(t)$ continues to increase, the further decrease of the systems energy cannot be proceeded by the decrease of its rest mass $m$. Namely, the particle in rest in
both worlds has total energy:
\begin{equation}
m c_{re} ^2 + mc_{im} ^2 =0.
\end{equation}
 This holds irrespective to the value of $m$. So, from the state with the vanishing energy at $t_a$, the system goes to the state with small negative kinetic energy of
the motion in the real world:
\begin{equation}
 -{mc_{re} ^2 \over (1- {v_{re} ^2 \over c_{re} ^2})^{1\over 2})} - mc_{im} ^2 .
\end{equation}
 This means that the mass of the system becomes negative. With the negative rest mass the fourth component of the 4-vector of momentum in the real world is inverted.
 And, if the positive fourth component of the 4-vector of momentum in the real world, which was the case before $t_a$, caused the propagation in the positive direction
 of the fourth component of the 4-vector of position in the real world $c_{re} t$, then the negative one causes propagation in the negative direction. As is well known,
 having negative mass is equivalent to the propagation backward in time, so at $t_a$, by acquiring the negative mass, the systems starts to propagate towards the past.
This holds for both the real and the imaginary world.

 On the other hand, the quantum system, like in the case of spatial barrier, can tunnel this temporal barrier. If before $t_1$ it was in the state $\vert p_{re} ^0
 \rangle$ and $ \vert p_{im} ^0 \rangle$, where $p_{im} ^0 =0$, then it can be found after $t_2$. As was said, at the moment that was the turning point for the
 classical system, the quantum one would not be present since, for all the time until it is observed, it would be in an eigenstate of energy. So to say, being in a
 state with sharp value of energy, it potentially exists before, during and after the period with the barrier, and the existence of the quantum system is not 
concentrated to any moment of time in particular - uncountably many moments of time are superposed in the actual state of the system, which is the state with sharp 
value of the energy. Its  $\vert p_{re} (t) \rangle$ and $ \vert p_{im} (t) \rangle$ would vary with respect to $t$ in such a way that the energy of the system plus 
$W(t)$ is equal to the  initial ${p_{re}^{0^2}\over 2m}$. Concretely, in the case when $W(t)$ is in the shape of the square between $t_a$ and $t_b$, with the value 
$W_0$, $W_0 >  {p_{re}^{0^2}\over 2m}$, the state of the quantum system before the barrier is:
\begin{equation}
 \int _{-\infty} ^{+\infty} \int _{-\infty} ^{+\infty} \int _{-\infty} ^{t_a} e^{{-1\over i \hbar} p_{re} ^0 \cdot q_{re}} e^{{-1\over i \hbar} p^0 _{im} \cdot q_{im}}
 e^{{1\over i \hbar} E_0 \cdot t } \vert q_{re} \rangle \otimes \vert q_{im} \rangle \otimes \vert t \rangle dq_{re} dq_{im} dt ,
\end{equation}
 where $p^0 _{im} = 0$ and $E_0 = {p_{re}^{0^2}\over 2m}$. Between the moments $t_a$ and $t_b$ the state is:
\begin{equation}
 \int _{-\infty} ^{+\infty} \int _{-\infty} ^{+\infty} \int _{t_a} ^{t_b} e^{{-1\over i \hbar} p^T _{re} \cdot q_{re}} e^{{-1\over i \hbar} p^T _{im} \cdot q_{im}}
 e^{{1\over i \hbar} E_T \cdot t } \vert q_{re} \rangle \otimes \vert q_{im} \rangle \otimes \vert t \rangle dq_{re} dq_{im} dt ,
\end{equation}
 where $p^T _{re}$, $p^T _{im}$ and $E_T$ are the tunneling momentas and energy, and where $p^T _{re}=0$, $p^T _{im}= ( {p_{re}^{0^2}\over 2m}- W_o)^{1\over2}$ and
$E_T ={p^{T^2} _{im} \over 2m}$. Finaly, after $t_b$ the state is:
\begin{equation}
 \int _{-\infty} ^{+\infty} \int _{-\infty} ^{+\infty} \int _{t_b} ^{+\infty} e^{{-1\over i \hbar} p^0 _{re} \cdot q_{re}} e^{{-1\over i \hbar} p^0 _{im} \cdot q_{im}}
 e^{{1\over i \hbar} E_0 \cdot t } \vert q_{re} \rangle \otimes \vert q_{im} \rangle \otimes \vert t \rangle dq_{re} dq_{im} dt .
\end{equation}

 In both cases, the spatial and the temporal barrier, the appropriate momentum of the classical system is reverted at the point where the system is confronted with the
 barrier. But, there is a strong difference between these two cases. Since there is no back in time, at the moment $t_a$ the classical system ceases to exist, it is no
 longer present in later times, and the force executed by the barrier is the cause of its death and destruction, so to say. (Dead means to be in rest.) In this context,
 for the quantum system it could be said that the tunneling is the process of dying at $t_a$ in the real world and immediate reincarnation in the imaginary world and,
at the moment $t_b$, resurrection in the real world.

\section{Black hole}

 When a black hole is looked from the center it looks like a barrier. The atractive gravitational force, directioned towards the center, plays the same role as the
 repulsive force of the potential barrier considered above. For the potential barrier it is important to know at which point the barrier starts and how high it can be.

 The particle in gravitational field, at rest at $r=+\infty$, has vanishing energy [33]. During its fall in a black hole, its velocity increases. At the distance 
$r^E$ from the  center, where $r^E$ is the Schwarzschild radius, the velocity of the particle becomes equal to the speed of light. Since the speed of light sets the 
uper bound for all velocities, the particle cannot move faster. This means that at the $r^E$ is the bottom of the potential well or, looked from the center of a black 
hole, at $r^E$ starts the potential barrier. The depth of the potential well is finite and it is equal to the value of the gravitational potential at $r^E$, which means that the
 potential is constant from $r=0$ to $r^E$. In other words, the gravitational force equals zero within the event horizon - the sphere of radius $r^E$, and the speed
 of light, as the finite uper bound for velocities, determines the depth of the potential well.

 After its pasage through the event horizon, the particle thermalises with the matter that has been already present in a black hole, so its velocity drops down in
 time. Due to the absence of the gravitational force within the sphere of radius $r^E$, matter is not shrinking to a single point and, therefore, there is no
 singularity at $r=0$ (all of the matter contained within the event horizon is not concentrated within the single point).

 If a classical system, with velocity $v$, $v < c$, that is directioned outwards a black hole, tries to escape it, it would be pulled back by attractive gravitational
 force. In the case with the potential barrier considered above, the repulsive force and insufficient energy were the reasons why the system could not go beyond the 
certain point, while here the attractive force, directioned towards the center of a black hole, and insufficient energy are the reasons why the system cannot escape. 
The quantum system also behaves like it behaved in the case of the potential barrier.

 In the case of the square spatial barrier the state of the quantum system is:
$$
 \int _{-\infty} ^{q_a} \int _{-\infty} ^{+\infty} \int _{-\infty} ^{+\infty} e^{{-1\over i \hbar} p_{re} ^0 \cdot q_{re}}e^{{-1\over i \hbar} p_{im} ^0 \cdot q_{im}}
 e^{{1\over i \hbar} E_0 \cdot t } \vert q_{re} \rangle \otimes \vert q_{im} \rangle \otimes \vert t \rangle dq_{re} dq_{im} dt +
$$
$$
 +\int _{q_a} ^{q_b} \int _{-\infty} ^{+\infty} \int _{-\infty} ^{+\infty} e^{{-1\over i \hbar} p_{re} ^T \cdot q_{re}} e^{-{1\over i\hbar} p_{im} ^T \cdot q_{im}}
 \cdot e^{{1\over i \hbar} E_0 \cdot t } \vert q_{re} \rangle \otimes \vert q_{im} \rangle \otimes \vert t \rangle dq_{re} dq_{im} dt +
$$
\begin{equation}
+\int _{q_b} ^{+\infty} \int _{-\infty} ^{+\infty} \int _{-\infty} ^{+\infty} e^{{-1\over i \hbar} p_{re} ^0 \cdot q_{re}} e^{{-1\over i \hbar} p_{im} ^0 \cdot q_{re}}
 e^{{1\over i \hbar} E_0 \cdot t } \vert q_{re} \rangle \otimes \vert q_{im} \rangle \otimes \vert t \rangle dq_{re} dq_{im} dt ,
\end{equation}
 where $p_{re} ^0$ is the initial real momentum, $p_{im} ^0 =0$, $p_{re} ^T =0$ and $p_{im} ^T = ( {p_{re}^{0^2}\over 2m}- V_o)^{1\over2}$. The quantum
system can escape from a black hole by tunneling the potential
barrier. Its state in this situation is:
$$
 \int _{0} ^{r_a} \int _{-\infty} ^{+\infty} \int _{-\infty} ^{+\infty} e^{{-1\over i \hbar} p_{re} ^0 \cdot r_{re}}e^{{-1\over i \hbar} p_{im} ^0 \cdot q_{im}}
 e^{{1\over i \hbar} E_0 \cdot t } \vert r_{re} \rangle \otimes \vert q_{im} \rangle \otimes \vert t \rangle dr_{re} dq_{im} dt +
$$
\begin{equation}
+\int _{r_a} ^{+\infty} \int _{-\infty} ^{+\infty} \int _{-\infty} ^{+\infty} e^{{-1\over i \hbar} p_{re} ^T \cdot r_{re}} e^{-{1\over i\hbar} p_{im} ^T \cdot q_{im}}
 \cdot e^{{1\over i \hbar} E_0 \cdot t } \vert r_{re} \rangle \otimes \vert q_{im} \rangle \otimes \vert t \rangle dr_{re} dq_{im} dt ,
\end{equation}

where $p_{re} ^0$ is the initial real $p_{im} ^0 =0$, $p_{re} ^T =0$ and $p_{im} ^T = (2m( V(r^E) + {p_{re}^{0^2}\over 2m} - V(r_{re}))^{1\over2}$ and $V(r_{re})= 
- \gamma  {m \cdot M \over r_{re} }$. So, due to the possibility to tunnel the barrier, the quantum system could be found out of the event horizon. For the systems 
that have evaporated from a black hole it is usually said that they are virtual particles. The attribute virtual is used in order to underline that their kinetic 
energies need not to be positive. Within the complexified quantum mechanics it becomes obvious that the evaporated particles are the standard particles that have 
the nonvanishing imaginary momentum, propagating forwards, not bekwards, in time (as is the case for some Feynman diagrams).

\section{Expansion of universe}

 Statement that the expansion of the universe is accelerating is followed by the question what causes such expansion. Some repulsive force, with negative pressure, is
 needed to explain acceleration of the expansion of the universe. The complexified classical mechanics offers the possibility to design a toy model of the universe, 
that is characterized with the accelerating expansion.

 Namely, if the universe is taken to be the hypersphere, then everything that exists in our world is within the real part of the hypersurface and the following
Hamilton function might be introduced:
\begin{equation}
 H ={{p^{T^2} _{re}} \over 2m } + {{p^{R^2} _{im}} \over 2m } - k { {q^{R^2} _{im}} \over q^T _{re} } ,
\end{equation}
 where $q^R _{im}$ and $p^R _{im}$ stand for the imaginary radius and corresponding momentum and $q^T _{re}$ and $p^T _{re}$ are some real tangential coordinate and
 momentum. Since we live in the real world, we can only observe some real coordinates and momenta. Within our world, let $q^T _{re}$ represent some coordinate that
 we are observing in order to conclude what is the rate of expansion of the universe. With the imaginary component $q^R _{im}$ we want to formalize impossibility to
measure the radius of the universe.

 Since ${q^{R^2} _{im}}$ is negative, for positive distance ${q^T _{re}}$ and positive imaginary radius ${q^R _{im}}$
the accelerations:
\begin{equation}
a^R _{re} = -{k\over m}{{q^{R^2} _{im}} \over {q^{T^2} _{re}}},
\end{equation}
 and
\begin{equation}
a^R _{im} = 2{k\over m} {q^R _{im} \over q^T _{re} },
\end{equation}
 are both positive and $v^T _{re} (t)= 2 {k\over m} t$ and $v^R _{im} (t)= i 2^{{3\over 2}} {k\over m} t$ both increase as the time lapses.

 The concept of dark energy is introduced with the purpose to explain acceleration of the expansion of the universe. With this toy model, where the potential is the
 product of the gravity like potential (real part) and the oscillator like potential (imaginary part), we wanted to show that, so to say, dark energy in the ordinary 
real world might be the ordinary energy in a dark world - the imaginary world. As are the positive energies natural for the real world, the negative energies are 
natural for the imaginary world.

\section{Conclusion}

The complexified quantum and classical mechanics were founded by the introduction of the anti self-adjoint operators of coordinate and momentum for the quantum 
mechanics and the imaginary variables of classical system. Within the proposed formalism, the tunnelings through the barriers, spatial and temporal, were thoroughly 
discussed. It is a well known fact that, during the tunneling through the potential barrier, the quantum system is characterized by the imaginary value of the momentum. 
This appears to be strange within the framework of standard quantum mechanics since only the self-adjoint operator of momentum is used, whose spectrum is real. The 
complexified quantum mechanics offered self consistent description of the tunneling through the spatial barrier where the unavoidable imaginary value of the momentum 
does not contradict the nature of the operator of imaginary momentum. The enlarged quantum mechanics was used in discussion of the tunneling through the temporal 
barrier and, in order to make the argumentation more transparent, the previously introduced operator of time formalism was employed. The complexified version of the 
classical mechanics, on the other hand, was used in the discussion regarding the systems passage through the event horizon, that was found to be the border of the 
potential barrier around a black hole. Finally, this new formalism was used with the purpose of designing a toy model of the expanding universe. It was shown that the 
acceleration of the expansion of the universe can be formaly described by useing the imaginary variables. 

The main conclusions are the following. The tunneling through the potential barrier shows that standard formulation of quantum mechanics, and consequently of 
classical mechanics as well, have to be enlarged. The quantum system can tunnel through the temporal barrier, while such barrier destroys the classical systems. There 
is no singularity at the center of a black hole and there is no need for dark energies since the negative energies are natural for the imaginary world.

\section{Acknowledgement}

 The author acknowledges funding provided by the Institute of Physics Belgrade, through the grant by the Ministry of Education, Science and Technological Development of
the Republic of Serbia.

\end{document}